\documentclass[11pt]{article}
\usepackage{amsfonts}
\usepackage{float}
\usepackage{amsmath}
\usepackage{booktabs}
\usepackage{mathtools}
\usepackage{color}
\usepackage{multibib}
\usepackage{bm}
\usepackage{xurl}
\usepackage{setspace}
\usepackage{algorithm}
\usepackage{authblk}  
\usepackage[nohead]{geometry}
\usepackage{algcompatible}
\usepackage[round]{natbib}
\usepackage{tikz}
\usetikzlibrary{arrows.meta, positioning}
\usepackage{calc}
\usepackage{hyperref}
\usepackage{multirow}
\usetikzlibrary{shapes.geometric, arrows, positioning,shapes,calc}
\usepackage{longtable}
\usepackage{arydshln}
\usepackage{caption} 

\DeclareMathOperator*{\minimize}{\text{minimize}}

\def\A{{\mathbf{A}}}
\def\B{{\mathbf{B}}}

\def\M{{\mathbf{M}}}

\def\X{{\mathbf{X}}}

\def\W{{\mathbf{W}}}
\def\I{{\mathbf{I}}}

\def\Y{{\mathbf{Y}}}
\def\Z{{\mathbf{Z}}}

\def\bbeta{{\bm{\beta}}}
\def\beps{{\bm{\epsilon}}}
\def\bmeta{{\bm{\eta}}}

\makeatletter
  \renewcommand*\env@matrix[1][*\c@MaxMatrixCols c]{%
    \hskip -\arraycolsep
    \let\@ifnextchar\new@ifnextchar
  \array{#1}}
\makeatother

\begin{document}
\title{CHIMA: a correlation-aware high-dimensional mediation analysis with its application to the living brain project study}

\author[1]{Samuel Osarfo}
\author[1]{Sangyoon Yi}
\author[2]{Weijia Fu}
\author[2]{Seungjun Ahn}

\affil[1]{Department of Statistics, Oklahoma State University, Stillwater, OK, 74078, USA}
\affil[2]{Department of Population Health Science and Policy, Icahn School of Medicine at Mount Sinai, New York, NY, 10028, USA}

\date{}

\maketitle
\sloppy%

\textbf{Abstract:} 
Mediation analysis examines the pathways through which mediators transmit the effect of an exposure to an outcome. In high-dimensional settings, the joint significance test is commonly applied using variable screening followed by statistical inference. However, when mediators are highly correlated, existing methods may experience reduced statistical power due to inaccurate screening and residual bias in asymptotic inference. To address these issues, we propose CHIMA (Correlation-aware High-dimensional Mediation Analysis), an extension of a recently developed high-dimensional mediation analysis framework that enhances performance under correlation by integrating two advances: (i) high-dimensional ordinary least squares projection for accurate screening under correlation; and (ii) approximate orthogonalization for bias reduction. Simulation studies demonstrate that CHIMA effectively identifies active mediators even in the presence of strong correlations and outperforms competing methods across various settings. We further apply CHIMA to ribonucleic acid sequencing (RNA-seq) from the Living Brain Project, identifying genes that mediate the effect of Parkinson’s disease on brain cell composition, thereby revealing cell-type-specific mechanisms of disease.

\strut \textbf{Keywords:} Correlated mediators, High-dimensional mediation analysis, Joint significance test, Living brain project, Parkinson's disease 

\section{Introduction}\label{sec:intro}
Mediation analysis, formalized by \cite{baron1986moderator}, investigates how an exposure affects an outcome through intermediate variables called mediators. Its ability to uncover causal mechanisms has made it widely applicable across several disciplines such as public health, genomics, and psychology \citep{lockwood2010mediation}. Recently, with the development of advanced data collection techniques, high-dimensional data become increasingly common in many areas of scientific research, which leads to increasing interest in developing methodology for high-dimensional mediation analysis. In our motivating data, we analyze RNA-seq data from the Living Brain Project to investigate molecular mechanisms underlying Parkinson’s disease. Previous studies reveal concurrent changes in gene expression and cell-type composition, suggesting a mediating role for gene expression. To explore this, we perform high-dimensional mediation analysis using preprocessed data on $58,929$ genes from $352$ cases and $164$ controls, aiming to identify genes that mediate the effect of disease status on cell-type composition.

Various methods have been developed to identify significant mediators in high-dimensional settings, see \cite{clark2023methods} for a review. A seminal work along the line of our work is HIMA \citep{zhang2016estimating}, which proceeds in three steps: (i) applying sure independence screening (SIS) by \cite{fan2008sure} to reduce dimensionality; (ii) computes each pair of $p$-values corresponding to the selected mediators in (i) from both mediator and outcome models using ordinary least squares and minimax concave penalty (MCP) by \cite{zhang2010nearly}; and (iii) performing a joint significance test based on the resulting $p$-values from (ii). Building on such framework, HDMA \citep{gao2019testing} replaces MCP with the debiased Lasso \citep{zhang2014confidence}, while HIMA2 \citep{perera2022hima2} further refines both the screening and testing steps using the method of \cite{dai2022multiple}.

Despite their widespread use, the aforementioned approaches would show low power when mediators are correlated, which is a common feature of RNA-seq data in our motivating dataset \citep{love2014moderated}. In particular, SIS could miss important variables since unimportant mediators highly correlated with active ones are likely to be selected. Additionally, HDMA and HIMA2 based on debiased Lasso require the remaining bias to vanish asymptotically for valid inference. In practice, however, such bias often exists, which can distort the sampling distribution and reduce inferential accuracy \citep{yi2022projection}. Since variables such as gene expressions are likely to be dependent in our high dimensional setup \citep{love2014moderated}, there is a clear need for a new method that remains robust under strong correlation among mediators.

To address such issue, we propose a method referred to as CHIMA, a correlation-aware high-dimensional mediation analysis. The main idea is to apply two recently developed methods in the first two steps of joint significance test. For screening, we apply the high-dimensional ordinary least squares projection (HOLP) by \cite{wang2016high}. Since the resulting rank from HOLP estimates is close to that of the true non-zero coefficients even under correlation, it satisfies the sure screening property without requiring the marginal correlation assumption of SIS. Then, we alternatively use the approximate orthogonalization by \cite{battey2023inference} instead of debiased Lasso. Similar to the BRP approach by \cite{yi2022projection}, it formulates optimization to directly minimize the resulting bias. Thus, as in \cite{yi2022projection}, it often shows a better finite sample performance compared to the debiased Lasso. Additionally, it admits a simple closed-form solution which lessens computational burden while the debiased Lasso requires choosing tuning parameters and the nodewise regression \citep{meinshausen2006high}.

The rest of this paper is organized as follows. In Section \ref{sec:method}, we introduce our method after briefly reviewing the existing frameworks in high-dimensional linear model. We compare the performance of each method in Section \ref{sec:sim} by conducting extensive simulation studies. We apply CHIMA for the real data set from living brain project in Section \ref{sec:real}. Conclusion and future research directions are discussed in Section \ref{sec:conc}. Our method can be implemented using the  \texttt{R} package \texttt{CHIMA} available in \url{https://github.com/samuelOsarfo/CHIMA}.

\section{Methods}\label{sec:method}

\subsection{Setup}
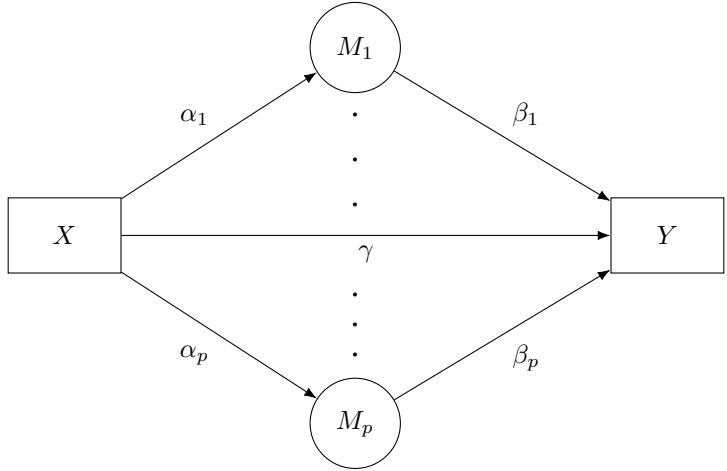
\begin{figure}[t]
\centering
\begin{tikzpicture}[
  node distance=2cm and 2.5cm,
  every node/.style={font=\small},
  mediator/.style={circle, draw, minimum size=1.2cm},
  main/.style={rectangle, draw, minimum height=1cm, minimum width=1.5cm},
  >=Latex
  ]

\node[main] (X) {$X$};
\node[mediator, right=2.5cm of X, yshift=2.5cm] (M1) {$M_1$};
\node[mediator, right=2.5cm of X, yshift=-2.5cm] (Mp) {$M_p$};
\node[main, right=6.5cm of X] (Y) {$Y$};

\node at ($(M1)!0.5!(Mp) + (0,1.6)$) {\Large $\cdot$};
\node at ($(M1)!0.5!(Mp) + (0,1.0)$) {\Large $\cdot$};
\node at ($(M1)!0.5!(Mp) + (0,0.4)$) {\Large $\cdot$};
\node at ($(M1)!0.5!(Mp) + (0,-0.8)$) {\Large $\cdot$};
\node at ($(M1)!0.5!(Mp) + (0,-1.2)$) {\Large $\cdot$};
\node at ($(M1)!0.5!(Mp) + (0,-1.6)$) {\Large $\cdot$};

\draw[->] (X) -- (M1) node[midway, above left] {$\alpha_1$};
\draw[->] (X) -- (Mp) node[midway, below left] {$\alpha_p$};

\draw[->] (M1) -- (Y) node[midway, above right] {$\beta_1$};
\draw[->] (Mp) -- (Y) node[midway, below right] {$\beta_p$};

\draw[->] (X) -- (Y) node[midway, below] {$\gamma$};

\end{tikzpicture}
\caption{A diagram of high-dimensional mediation model with effect parameters: exposure ($X$), $p$ mediators ($M_1,\ldots,M_p$), and outcome ($Y$).}
\label{fig1}
\end{figure}

Let $X$ be an exposure or treatment, $Y$ be an outcome, and $M_{j}$ be the $j$-th potential mediator for $j=1,\dots,p$. We seek to identify which variable $M_{j}$ mediates the effect of $X$ on $Y$. For high-dimensional mediation analysis, it is common to formulate such mediation effects by using the following linear structural equation models:
\begin{equation}\label{lmod}
\begin{aligned}
\text{(Mediator model)}\quad &M_{j} = \alpha_{j} X + u_{j},  \\
\text{(Outcome model)}\quad &Y = \sum_{j=1}^{p}\beta_{j} M_{j} + \gamma X + \epsilon,
\end{aligned}    
\end{equation}
where $\alpha_{j}$ is the effect of exposure on $j$-th mediator, $\beta_{j}$ is the effect of $j$-th mediator on outcome adjusted for $X$ and other mediators and $\gamma$ is the direct effect of $X$ on $Y$ adjusted for each $M_{j}$; and $u_{j}$ and $\epsilon$ are i.i.d. random errors with zero mean and common variances $\sigma_{u}^{2}$ and $\sigma_{\epsilon}^{2}$ in each model, respectively. We omit intercept and auxiliary covariates in \eqref{lmod} for simplicity, but our method can easily be extended to include them. Figure \ref{fig1} describes such underlying mechanism under \eqref{lmod} between exposure $X$, each mediator $M_{j}$ and outcome $Y$. The global indirect or mediation effect of $X$ on $Y$ through $M_{j}$'s is  $\sum_{j=1}^{p}\alpha_{j}\beta_{j}$ and the total effect of $X$ on $Y$ is $\gamma+\sum_{j=1}^{p}\alpha_{j}\beta_{j}$. Depending on one's interest, we would be interested in not only $\sum_{j=1}^{p}\alpha_{j}\beta_{j}$ as a whole but also the individual product terms $\alpha_{j}\beta_{j}$. Such product of two effects is referred to be the mediation contributions of the $j$-th mediator and the main goal of this paper is to identify which $M_{j}$ has non-zero mediation contribution, i.e., $\alpha_{j}\beta_{j}\neq0$. 

A few noteworthy features of the data that we consider in this work are (i) high-dimensionality; (ii) presence of correlation among mediators. As for (i), the total number of mediators $p$ is often much larger than the available sample size $n$, which is common nowadays due to the technological advancement. Such high-dimensionality prevents from using the ordinary least square (OLS) estimates in the outcome model. Thus, it is often considered to be reasonable to assume the sparsity of either $\alpha_{j}$ or $\beta_{j}$. On the other hand, (ii) is also common especially when variables from omics data are considered as mediators. Though it is also possible to consider the potential mediators one-at-a-time in separate outcome models, such marginal approach can be problematic when the mediators are correlated \citep{clark2023methods}. Thus, we evaluate the mediators jointly by fitting an outcome model that adjusts the effect of each mediator for the others as in \eqref{lmod}. Both (i) and (ii) are relevant to the real data considered in this paper because the total number of genes is much larger than the available sample size and those genes from RNA-seq data tend to be highly correlated \citep{love2014moderated}.

\subsection{Review of Existing Approaches}

Various methods for high-dimensional mediation analysis have been proposed based on penalized regression, dimension reduction to name a few. We refer to \cite{clark2023methods} for a comprehensive review. Among them, several penalized regression approaches are the lines of study that are closest to our approach such as HIMA \citep{zhang2016estimating}, HDMA \citep{gao2019testing} and HIMA2 \citep{perera2022hima2}. They share the key idea which consists of the following three steps: 

\begin{enumerate}
    \item[]\textbf{Step 1.} Conduct variable screening to choose candidate of significant mediators
    \item[]\textbf{Step 2.} Consider the reduced outcome model with the selected mediators in Step 1 and compute two $p$-values from that model to test each null hypothesis $\alpha_{j} = 0$ and $\beta_{j}=0$ for each chosen mediators, respectively
    \item[]\textbf{Step 3.} Conduct testing $H_{0j}:\alpha_{j}\beta_{j}=0$ by using both $p$-values from Step 2
\end{enumerate}

In Step 1, each aforementioned method uses SIS \citep{fan2008sure} or its variant for mediator selection. In Step 2, each $p$-value $p_{\alpha,j}$ can be computed with the corresponding OLS estimates for $\alpha_{j}$. HIMA considers outcome model with the selected mediators only and computes $p$-value $p_{\beta,j}$ using MCP \citep{zhang2010nearly}. Alternatively, both HDMA and HIMA2 use debiased Lasso \citep{zhang2014confidence} to compute asymptotically valid $p$-value to test $\beta_{j}=0$. In Step 3, the joint significance test using the maximum of $p_{\alpha,j}$ and $p_{\beta,j}$ is conducted for both HIMA and HMDA. However, such test is known to be overly conservative especially when both $\alpha_{j}=0$ and $\beta_{j}=0$, which is often believed to be the case under high-dimensional regime \citep{du2023methods}. To address such issue, HIMA2 instead adopts the HDMT \citep{dai2022multiple} based on the mixture as reference distribution in an attempt to take into account the composite nature of testing $H_{0}:\alpha_{j}\beta_{j}=0$.  

\subsection{Proposed Method}
This work is motivated by the two aspects from Steps 1-2 of the existing methods. While SIS can enjoy the so-called sure screening property under suitable conditions, such accuracy is achieved only when the active mediators have large marginal correlations with the response. However, such underlying assumption rarely holds in practice when many predictors are highly correlated in high-dimensional regime \citep{wang2016high}. And this would also be the case in high-dimensional mediation analysis as omics-type mediators are often correlated to each other. Thus, several existing methods based on SIS would miss some important mediators with non-zero effects. Additionally, debiased Lasso could show undesirable empirical performance, that is, the average coverage rate for non-zero effect could be far lower than the nominal levels in finite sample \citep{yi2022projection}. Such phenomenon can be attributed to that the remaining bias still exists thus cannot be negligible. Thus, the asymptotic normality of debiased Lasso estimator would not be accurate in practice. To address such challenges, we use each alternative approach for Steps 1-2 described in the following:

\begin{itemize}

    \item[]\textbf{Proposed Step 1.} Instead of the SIS, we propose to apply HOLP by \cite{wang2016high} for mediator screening. Under the outcome model \eqref{lmod}, the HOLP estimates can be given as
    \begin{equation}\label{holp}
    [\tilde{\beta}_{1}\ \tilde{\beta}_{2}\ \dots\ \tilde{\beta}_{p}\ \tilde{\gamma}]^\top= \Z^\top ( \Z \Z^\top)^{-1}\Y    
    \end{equation}
    where $\Z=[\M_{1}\ \M_{2}\ \dots\ \M_{p}\ \X]\in\mathbb{R}^{n \times (p + 1)}$, $\M_{j}$, $\X$ and $\Y$ are $n$-dimensional vector of $j$-th mediator, exposure and response, respectively, for $j=1,\dots,p$. The intuition behind \eqref{holp} is as follows: since our goal is to screen out the mediators with $\beta_{j}\neq0$, it is sufficient to consider the linear estimates $\A\Y$ for some $\A\in\mathbb{R}^{(p+1)\times n}$ that preserves the order of true $|\beta_{j}|$. Multiplying $\A$ on both sides of \eqref{lmod} gives us 
    \[
    \A\Y = \A\Z[\bbeta\ \gamma]^\top + \A\beps,
    \]
    where $\bbeta=[\beta_{1}\ \dots\ \beta_{p}]^\top\in\mathbb{R}^{p\times1}$ and $\beps\in\mathbb{R}^{n\times1}$ is the vector of random error $\epsilon_{i}$ associated with the $i$-th sample. Note that $\A\Y \approx \A\Z[\bbeta\ \gamma]^\top $ since $E[\epsilon_{i}]=0$. This implies that the least square estimates $(\Z^\top \Z)^{-1}\Z^\top\Y$ would be a reasonable estimates (i.e., $\A=(\Z^\top \Z)^{-1}\Z^\top$) that enables to exactly recover of the true $\beta$ since $\A\Z$ would be an identity matrix in such case. However, since $(\Z^\top \Z)^{-1}$ does not exist for $p>n$, \cite{wang2016high} introduced a Moore-Penrose inverse $\Z^\top ( \Z\Z^\top)^{-1}$ for $\A$ instead. This choice leads the resulting $\A\Z$ to be diagonally dominant under suitable conditions, which helps to preserve the order of $|\beta_{j}|$ and select important variables. Theoretically, the HOLP estimator \eqref{rholp} can be shown to enjoy the sure screening property without the marginal correlation assumption as required by SIS.

    \item[]\textbf{Proposed Step 2.} To compute the $p$-value for $H_{0}:\beta_{j}=0$ in Step 2, we apply the approximate orthogonalization (AO) approach \citep{battey2023inference} instead of debiased Lasso. Using the matrix notation, we motivate this by rewriting \eqref{lmod} as
    \begin{equation}\label{redmod}
    \Y = \M_{j}\beta_{j} + \M_{-j}\bbeta_{-j} + \X \gamma+ \beps, 
    \end{equation}
    where $\M_{-j}\in\mathbb{R}^{n \times (p-1)}$ and $\bbeta_{-j}\in\mathbb{R}^{(p-1) \times 1}$ are matrix and vector after removing the $j$-th column and element from $\M$ and $\bbeta$, respectively. Its key idea is to introduce $\B_{j}\in\mathbb{R}^{n\times n}$ which orthogonalizes $\B_{j}\M_{j}\in\mathbb{R}^{n\times 1}$ to each column of $\B_{j}[\M_{-j}\ \X]$ as much as possible. To understand the reason, note that multiplying $v_{j}^\top=\M_{j}^\top\B_{j}^\top\B_{j}\in\mathbb{R}^{1\times n}$ on both sides of \eqref{redmod} gives 
    \begin{equation}\label{redmod2}
    \begin{aligned}
    v_{j}^\top\Y &= v_{j}^\top\M_{j}\beta_{j} + v_{j}^\top \M_{-j}\bbeta_{-j} + v_{j}^\top\X\gamma + v_{j}^\top\beps \\
    &= v_{j}^\top\M_{j}\beta_{j} + v_{j}^\top\W_{-j}\bmeta_{-j} + v_{j}^\top\beps \\
    \end{aligned}
    \end{equation}
    where $\W_{-j}=[\M_{-j}\ \X]$ and $\bmeta_{-j}=[\bbeta_{-j}^\top\ \gamma]^\top$. Dividing both sides of \eqref{redmod2} by $v_{j}^\top\M_{j}\in\mathbb{R}$, we have
    \begin{equation}\label{redmod3}
    (v_{j}^\top \M_{j})^{-1} v_{j}^\top \Y 
    = \beta_{j} + \underbrace{(v_{j}^\top \M_{j})^{-1}v_{j}^\top \W_{-j}\bmeta_{-j}}_{:=R_{j}} +(v_{j}^\top \M_{j})^{-1} v_{j}^\top \beps.     
    \end{equation}
    The last term $(v_{j}^\top \M_{j})^{-1} v_{j}^\top \beps$ on the RHS of \eqref{redmod3} can be shown to converge to a Normal distribution under some suitable assumptions on $\epsilon_{i}$. This implies that $(v_{j}^\top \M_{j})^{-1} v_{j}^\top \Y$ on the LHS of \eqref{redmod3} asymptotically follows the Normal distribution as well if the bias $R_{j}=0$. As a result, we can compute the $p$-value under $H_{0}:\beta_{j}=0$.
     
    Unfortunately, no such $v_{j}$ exists that satisfies $R_{j}=0$ through exact orthogonality, i.e., $v_{j}^\top\W_{-j}=\mathbf{0}$, under high-dimensional setting due to the Rank-nullity theorem \citep{axler2024linear}. However, we can still find $v_{j}$ approximately orthogonal to each column of $\W_{-j}$ so that $R_{j}\approx0$, which enables to conduct an inference for $\beta_{j}$ in asymptotic sense, as in debiased Lasso. To find such $v_{j}$, \cite{battey2023inference} formulate the following optimization problem:   
    \begin{equation}\label{br:opt}
    \minimize_{v_{j}} (v_{j}^\top \M_{j})^{-2} v_{j}^\top (\W_{-j}\W_{-j}^\top + \delta \I)v_{j},    
    \end{equation}
    where $\delta>0$ controls the overall contribution of the second term to the objective. \eqref{br:opt} can be motivated as follows: (i) the first term is equivalent to $\|(v_{j}^\top \M_{j})^{-1}v_{j}^\top \W_{-j}\|_{2}^{2}$ as an upper bound of the bias $R_{j}$ by the Cauchy-Schwarz inequality; (ii) $(v_{j}^\top \M_{j})^{-2}v_{j}^\top v_{j}$ in the second term is proportional to the conditional variance of $(v_{j}^\top \M_{j})^{-1} v_{j}^\top \beps$ given $v_{j}$. Thus, the optimization \eqref{br:opt} aims to minimize both the upper bound of bias and term relative to that to achieve efficiency. Under suitable conditions, it can be shown that a closed form solution $\hat{v}_{j}$ exists to \eqref{br:opt} such that 
    \begin{equation}\label{sol:proj}
    \hat{v}_{j} = \delta(\W_{-j}\W_{-j}^\top + \delta\mathbf{I})^{-1}\M_{j}.    
    \end{equation}
    Thus, we have the test statistic $\hat{\beta}_{j}=(\hat{v}_{j}^\top \M_{j})^{-1} \hat{v}_{j}^\top \Y$ that asymptotically follows Normal distribution with mean zero and variance $(\hat{v}_{j}^\top \M_{j})^{-2}(\hat{v}_{j}^\top\hat{v}_{j})\sigma_{\epsilon}^{2}$ as proven in \cite{battey2023inference} under suitable conditions. 
\end{itemize}

Combining the above, we summarize our proposed procedure in Algorithm \ref{algo}. For screening, the estimator \eqref{rholp} is also proposed in \cite{wang2016high}. It provides greater flexibility, as \eqref{rholp} gets close to the marginal screening operator $\Z^\top\Y$ used in SIS as $k\rightarrow\infty$, while reducing to \eqref{holp} when $k=0$. Though \cite{wang2016high} mention \eqref{rholp} performs similarly to \eqref{holp} in practice, we present \eqref{rholp} due to its more general form. In our implementation of CHIMA, we set $d=\lceil n/\log n \rceil$, $k=1$ and $\delta=1$ in Steps 1 and 2 of Algorithm \ref{algo}, respectively, where $\lceil \cdot \rceil$ denotes the rounded up integer. We observe that our results remain similar across a wide range of these tuning parameters, which is consistent with the robustness reported in \cite{wang2016high} and \cite{battey2023inference}. Lastly, we note that CHIMA is computationally more efficient than debiased LASSO, which requires computation for tuning parameter selection and nodewise regression \citep{meinshausen2006high} to estimate projection direction.

\begin{algorithm}[!t]
\begin{algorithmic}
\caption{CHIMA}\label{algo}
\STATE \textbf{Input}: $\X\in\mathbb{R}^{n\times 1}$, $\M\in\mathbb{R}^{n \times p}$, $\Y\in\mathbb{R}^{n\times 1}$, $d,k,\delta>0$, $\alpha\in(0,1)$
\STATE $\boldsymbol{\cdot}$ Step 1: Compute the Ridge-HOLP (RHOLP) estimates such that 
\begin{equation}\label{rholp}
[\tilde{\beta}_{1}\ \tilde{\beta}_{2}\ \dots\ \tilde{\beta}_{p}\ \tilde{\gamma}]^\top = \Z^\top(k\mathbf{I}_{n} + \Z \Z^\top)^{-1}\Y,    
\end{equation}
\hspace{1.5cm} where $\Z=[\M\ \X]$ and the OLS estimates $\hat{\alpha}_{j}$ for $j=1,\dots,p$.\\ \hspace{1.5cm} The set of potentially active mediators is chosen as
\[
\mathcal{S}=\left\{1\leq j \leq p:\text{the $j$-th mediator is among the top $d$ largest $|\hat{\alpha}_{j}\tilde{\beta}_{j}|$}\right\}.
\]

\STATE $\boldsymbol{\cdot}$ Step 2: For each $j\in\mathcal{S}$, compute the pair of $p$-values 
\[
\begin{aligned}
p_{\alpha,j}= 2\{1 - \Phi(\hat{\alpha}_{j}/\text{se}_{\alpha,j})\},\quad
p_{\beta,j}= 2\{1 - \Phi(\hat{\beta}_{j}/\text{se}_{\beta,j})\},
\end{aligned}
\]
\hspace{1.5cm} where 
\[
\begin{aligned}
&\text{se}_{\alpha,j}=\sqrt{(\X^\top\X)^{-1}\hat{\sigma}_{u}^{2}},\quad \hat{\sigma}_{u}^{2}=\widehat{\text{Var}}(u_{j}),\\
&\hat{\beta}_{j}=(\hat{v}_{j}^\top \M_{j})^{-1}\hat{v}_{j}^\top\Y, \quad \text{se}_{\beta,j}=\sqrt{(\hat{v}_{j}^\top \M_{j})^{-2}(\hat{v}_{j}^\top\hat{v}_{j})\hat{\sigma}_{\epsilon}^{2}},\quad\hat{\sigma}_{\epsilon}^{2}=\widehat{\text{Var}}(\epsilon),  \\  
\end{aligned}
\]  
\hspace{1.5cm} for $\hat{v}_{j}$ is the projection vector as in \eqref{sol:proj}.
\STATE $\boldsymbol{\cdot}$ Step 3: Estimate FDR as 
\[
\widehat{\text{FDR}}(t)=\dfrac{\hat{\pi}_{01}t + \hat{\pi}_{10}t + \hat{\pi}_{00}t^2}{\text{max}[1, R(t)]/|\mathcal{S}|}\quad \text{for}\quad t\in[0,1],
\]
\hspace{1.5cm} where $\hat{\pi}_{01},\hat{\pi}_{10}$ and $\hat{\pi}_{00}$ are the estimated proportions of the three disjoint null \\
\hspace{1.5cm} hypotheses, respectively, and $R(t)$ is the number of discoveries. Each quantity \\
\hspace{1.5cm} can be computed via \texttt{R} package \texttt{HDMT} given $\{(p_{\alpha,j},p_{\beta,j})\}_{j\in\mathcal{S}}$ from Step 2. \\
\hspace{1.5cm} For more details, we refer to Appendix \ref{app:step3}.

\STATE \textbf{Return}: the set of significant mediators such that 
\[
\mathcal{D}=\left\{j:\text{max}(p_{\alpha,j},p_{\beta,j}) \leq \hat{t}_{\alpha}\ \text{and}\ j\in\mathcal{S}\right\},
\]
\hspace{1.5cm} where $\hat{t}_{\alpha}=\text{sup}\{t\in[0,1]:\widehat{\text{FDR}}(t)\leq\alpha\}$.
\end{algorithmic}
\end{algorithm}

\section{Simulation studies}\label{sec:sim}
We conduct extensive simulation studies and compare CHIMA to HIMA, HIMA2 and HDMA. For the last three competing methods, we set $\lceil 2n/\log n \rceil$ as the number of screened mediators as suggested in the original papers. To illustrate the performance of AO against debiased Lasso, we further consider HOLP$+$DB, which uses debiased Lasso in Step 2 of Algorithm \ref{algo}. To assess the performance of each method, we consider three metrics: (i) screening rate; (ii) test power; and (iii) false discovery proportion (FDP). Note that HIMA2 applies SIS using the product of estimated $\alpha_{j}$ and $\beta_{j}$ from marginal models while both HIMA and HDMA employ SIS based on $\hat{\alpha}_{j}$ from \eqref{lmod}. Thus, HOLP is compared with HIMA2 and HIMA for (i).

For data generation, we consider each combination of $p=6000,8000$ and total number of active mediators $s_{11}=4,6$ with sample size $n=400$. The first $1.5\times s_{11}$ elements of $\alpha_{j}$ are set to be non-zero, while the rest are $0$. For $\beta_{j}$, we assign the first $s_{11}$ elements to be non-zero, followed by $s_{11}/2$ zeros, and another $s_{11}/2$ non-zero elements, with all remaining elements set to 0. Each non-zero coefficient is independently drawn from $\text{Unif}(0.3, 1)$ and randomly assigned to be either positive or negative. We set $\gamma=0.5$ and independently generate each element of $X$ and $\epsilon$ in the outcome model from $N(0,1.5)$ and $N(0,1)$, respectively. To incorporate various structures of correlation among mediators, we simulate correlated random errors $u_{i}=[u_{i1}\ \dots\ u_{ip}]^\top$ where $u_{ij}$ is the random error in \eqref{lmod} associated with the $j$-th mediator for the $i$-th subject. Similar to \cite{perera2022hima2}, each $u_{i}$ is independently simulated from $N_{p}(\mathbf{0},\mathbf{\Sigma})$ where covariance matrix $\mathbf{\Sigma}$ is of either compound symmetry (CS) with correlation $\rho=0.85$ or Toeplitz (TP) with  $\rho=\pm0.85$, both having unit diagonal entries. Additionally, motivated by \cite{wang2016high} and \cite{meinshausen2010stability}, we also consider the factor model for $u_{ij}$ such that 
\begin{equation}\label{mod:factor}
u_{ij}=\sum_{l=1}^{r}f_{il}\lambda_{lj} + \eta_{ij},
\end{equation}
where $f_{il}, \lambda_{lj} \stackrel{i.i.d.}{\sim} N(0,1)$ and $\eta_{ij} \stackrel{i.i.d.}{\sim} N(0,\tau^{2})$ with $r=2,4$ and $\tau=0.8$. Based on these settings, mediators and outcomes are generated according to \eqref{lmod} and $500$ independent simulation runs are performed for each scenario. To avoid redundancy, we report results for $p = 8000$ only, since similar patterns appear across other settings and with an alternative scheme with $\text{Unif}(0.5, 1.0)$ for non-zero coefficients; the corresponding results are included in Appendix \ref{app:sim}.

\renewcommand{\arraystretch}{1.2}
\begin{table}[!t]
\centering
\begin{tabular}{ccccc}
\hline
Covariance structure & $s_{11}$ & \multicolumn{3}{c}{Method} \\
& & HOLP & HIMA2 & HIMA \& HDMA \\
\hline
\multirow{2}{*}{Compound symmetry} & 4 & 0.9905 & 1.0000 & 1.0000 \\
\cline{2-5}
& 6 & 0.9893 & 1.0000 & 1.0000 \\
\hline
\multirow{2}{*}{Toeplitz ($\rho=0.85$)} & 4 & 0.9990 & 1.0000 & 1.0000 \\
\cline{2-5}
& 6 & 0.9983 & 1.0000 & 1.0000 \\
\hline
\multirow{2}{*}{Toeplitz ($\rho=-0.85$)} & 4 & 0.9835 & 0.9715 & 0.7805 \\
\cline{2-5}
& 6 & 0.9773 & 0.9787 & 0.7977 \\
\hline
\multirow{2}{*}{Factor with $r = 2$} & 4 & 0.9590 & 0.8965 & 0.3355 \\
\cline{2-5}
& 6 & 0.9557 & 0.8907 & 0.2863 \\
\hline
\multirow{2}{*}{Factor with $r = 4$} & 4 & 0.9455 & 0.8380 & 0.1840 \\
\cline{2-5}
& 6 & 0.9470 & 0.8283 & 0.1373 \\
\hline
\end{tabular}
\caption{Average screening rate of each method for each different covariance structures and $s_{11}=4,6$ at significance level $\alpha=0.05$ when $n = 400$ and $p=8000$}\label{tab1:screen}
\end{table}

Tables \ref{tab1:screen} shows the average screening rate across 500 simulations for each method. All of three similarly performs great though HOLP shows slightly lower screening rate for the first two covariance structures. For TP with negative correlation, HOLP and HIMA2 show similar screening rate while both HIMA and HDMA miss active mediators. For the factor model \eqref{mod:factor}, though the rate of each method gets decreased, HOLP still shows more accurate screening performance compared to the others. Thus, given that HOLP selects $\lceil n / \log n \rceil$ mediators—about half as many as the other methods—it is evident that HOLP generally outperforms the existing SIS-based approaches.

\renewcommand{\arraystretch}{1.2}
\begin{table}[!t]
\centering
\begin{tabular}{ccccccc}
\hline
Covariance structure & $s_{11}$ & \multicolumn{5}{c}{Method} \\
& & CHIMA & HOLP$+$DB & HIMA2 & HIMA & HDMA \\
\hline
\multirow{2}{*}{Compound symmetry} & 4 & 0.7375 & 0.8065 & 0.7895 & 0.7095 & 0.7765 \\
\cline{2-7}
& 6 & 0.7590 & 0.8077 & 0.7780 & 0.7307 & 0.7850 \\
\hline
\multirow{2}{*}{Toeplitz ($\rho=0.85$)} & 4 & 0.8670 & 0.9500 & 0.9455 & 0.7835 & 0.9860 \\
\cline{2-7}
& 6 & 0.8610 & 0.9680 & 0.9657 & 0.7707 & 0.9903 \\
\hline
\multirow{2}{*}{Toeplitz ($\rho=-0.85$)} & 4 & 0.7545 & 0.7140 & 0.6635 & 0.3880 & 0.5975  \\
\cline{2-7}
& 6 & 0.7803 & 0.7117 & 0.6570 & 0.4157 & 0.6033 \\
\hline
\multirow{2}{*}{Factor with $r = 2$} & 4 & 0.9165 & 0.9290 & 0.8485 & 0.3205 & 0.3295 \\
\cline{2-7}
& 6 & 0.9080 & 0.9180 & 0.8370 & 0.2693 & 0.2770 \\
\hline
\multirow{2}{*}{Factor with $r = 4$} & 4 & 0.8920 & 0.8955 & 0.7710 & 0.1745 & 0.1800 \\
\cline{2-7}
& 6 & 0.8850 & 0.8880 & 0.7480 & 0.1277 & 0.1347 \\
\hline
\end{tabular}
\caption{Average power of each method for each different covariance structures and $s_{11}=4,6$ at significance level $\alpha=0.05$ when $n = 400$ and $p=8000$}\label{tab1:power}
\end{table}

\begin{table}[!t]
\centering
\begin{tabular}{ccccccc}
\hline
Covariance structure & $s_{11}$ & \multicolumn{5}{c}{Method} \\
& & CHIMA & HOLP$+$DB & HIMA2 & HIMA & HDMA \\
\hline
\multirow{2}{*}{Compound symmetry} & 4 & 0.0305 & 0.0381 & 0.0037 & 0.0278 & 0.1839 \\
\cline{2-7}
& 6 & 0.0364 & 0.0404 & 0.0057 & 0.0104 & 0.1392 \\
\hline
\multirow{2}{*}{Toeplitz ($\rho=0.85$)} & 4 & 0.0521 & 0.0983 & 0.0788 & 0.0195 & 0.1685 \\
\cline{2-7}
& 6 & 0.0470 & 0.0745 & 0.0598 & 0.0163 & 0.1299 \\
\hline
\multirow{2}{*}{Toeplitz ($\rho=-0.85$)} & 4 & 0.0514 & 0.0945 & 0.0601 & 0.0442 & 0.3303 \\
\cline{2-7}
& 6 & 0.0729 & 0.1275 & 0.0793 & 0.0460 & 0.2557 \\
\hline
\multirow{2}{*}{Factor with $r = 2$} & 4 & 0.0609 & 0.0945 & 0.0114 & 0.0182 & 0.1000 \\
\cline{2-7}
& 6 & 0.0568 & 0.0944 & 0.0129 & 0.0160 & 0.0935 \\
\hline
\multirow{2}{*}{Factor with $r = 4$} & 4 & 0.0530 & 0.0958 & 0.0140 & 0.0232 & 0.1246 \\
\cline{2-7}
& 6 & 0.0636 & 0.1006 & 0.0123 & 0.0270 & 0.1230 \\
\hline
\end{tabular}
\caption{Average FDP of each method for each different covariance structures and $s_{11}=4,6$ at significance level $\alpha=0.05$ when $n = 400$ and $p=8000$}\label{tab1:fdp}
\end{table}

Tables \ref{tab1:power}-\ref{tab1:fdp} summarize the average power and FDP for each method. Both CHIMA and HIMA2 generally achieve higher power compared to HIMA and HDMA, while maintaining FDR control around the nominal level $0.05$ across all covariance structures. In contrast, HIMA tends to be overly conservative, particularly under TP with $\rho=-0.85$ and factor model. Although HDMA shows power comparable to HIMA2 under the CS and TP structures, its lack of multiple testing adjustment leads to inadequate control of the FDP, as noted in \cite{perera2022hima2}. For CS and TP with $\rho = 0.85$, HIMA2 slightly outperforms CHIMA in terms of power with differences up to $0.1$. However, CHIMA shows higher power in all other covariance settings. For instance, such gap is as large as $0.137$ when $r=4$ and $s_{11}=6$ under the factor model. This pattern also holds for $p=6000$ and under the alternative scheme for generating non-zero coefficients as shown in Appendix \ref{app:sim}. As shown in Table \ref{tab1:screen}, such phenomena can be attributed to the better screening performance of HOLP compared to that of HIMA2, which has been already noted in \cite{wang2016high}. Also, the effectiveness of AO can be supported by comparing CHIMA to HOLP$+$DB. Though HOLP$+$DB shows higher power than CHIMA, it fails to control FDR at the nominal level due to the residual bias in debiased Lasso, which results in an inaccurate sampling distribution. This highlights the additional benefit of CHIMA’s use of AO. Overall, CHIMA consistently demonstrates the most favorable performance, offering high power while effectively controlling FDR across our simulation settings. 

\section{Application study}\label{sec:real}
\subsection{Living Brain Project Study Data}
We are motivated by Parkinson's disease (PD), a progressive neurodegenerative disorder marked by selective dopaminergic neuronal loss and reactive gliosis in cortical and subcortical regions, yet the molecular mechanisms linking disease status to shifts in cellular composition remain understudied \citep{PDintro}. PD pathology is characterized by significant alterations in brain cell‐type composition, most notably neuronal loss, astrocytic gliosis (AST), oligodendrocyte (ODC) changes, and microglial (MG) activation \citep{PDpathology}. Recent RNA‐sequencing studies \citep{PDrnaseq1,PDrnaseq2, PDrnaseq3} have observed cell‐type‐specific gene expression changes in PD brains, concurrent with shifts in cell‐type proportions. These observations suggest that altered gene expression may contribute to changes in cellular composition, supporting the hypothesis that gene expression could serve as a mediator between disease status and cell‐type shifts. Although causality cannot be definitively established through observational mediation analyses, we use it as an exploratory tool to pinpoint genes whose expression changes serve as plausible intermediaries between PD status and cell type composition shifts, thereby generating hypotheses for downstream validation.

In this paper, preprocessed RNA-Seq data and clinical covariates from the Living Brain Project (LBP) \citep{LBPstudy} were obtained via the Mount Sinai Data Ark (\url{https://labs.icahn.mssm.edu/minervalab/resources/data-ark/lbp/}) and analyzed on the Minerva Supercomputer at the Icahn School of Medicine at Mount Sinai. The LBP is a state-of-the-art study that utilizes fresh brain tissue from living human donors, providing a novel opportunity to study brain biology rather than relying on postmortem tissue as a proxy, which remains the primary source of data in most neurological research. This dataset represents a distinctive resource for examining the molecular basis of human brain health and disease in living donors, offering insights beyond those achievable from postmortem tissues.

The sample size consisted of 516 participants, including $n$ = 352 with PD and $n$ = 164 control, with gene expression measured across 58,929 genes. Genes with zero expression across all samples were excluded, yielding 57,630 retained genes for the analysis. Gene expression values were subsequently normalized using the \texttt{DESeq2 R} package. The estimated cell-type fractions were ODC, neuronal (calculated as the sum of estimated glutamatergic and GABA-ergic neuronal cells), AST, and MG. In this causal mediation analysis, continuous age and binary postmortem status (postmortem vs. living brain samples) were regressed as covariates. 

\subsection{Results}

\begin{figure}[t]
\centering
\includegraphics[scale=0.65]{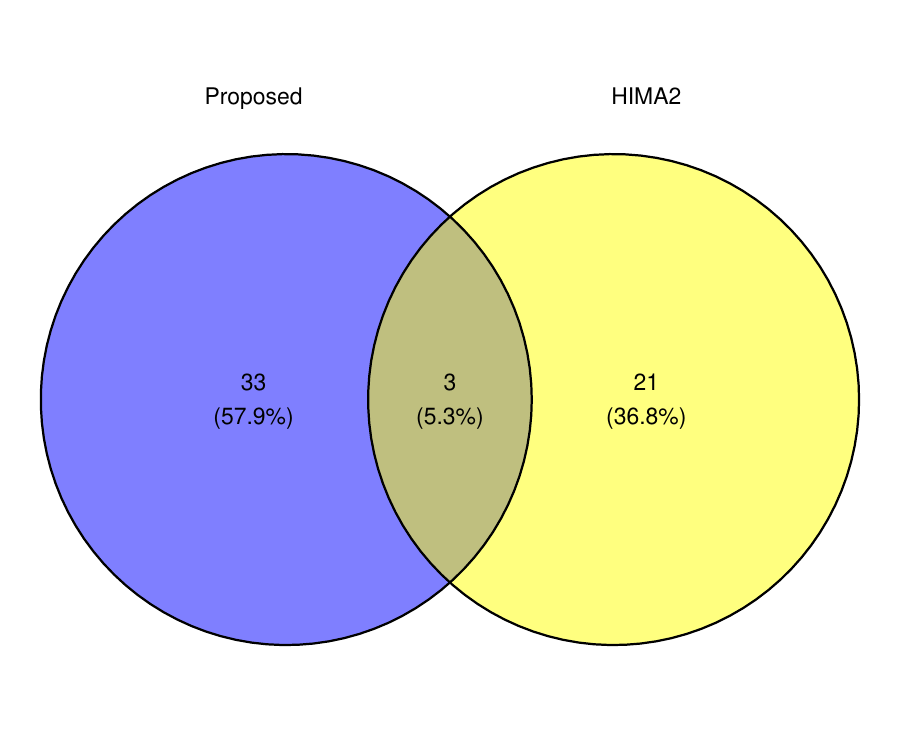}
\caption{Venn diagram illustrating the overlap of significant mediating genes identified by two methods: CHIMA (denoted as Proposed) and HIMA2, when modeling oligodendrocyte (ODC) estimates as the outcome. The diagram highlights both shared and method-specific mediators, providing a comparative assessment of mediator selection consistency across approaches.}
\label{LBPfig1}
\end{figure}

\begin{figure}[H]
\centering
\includegraphics[scale=0.65]{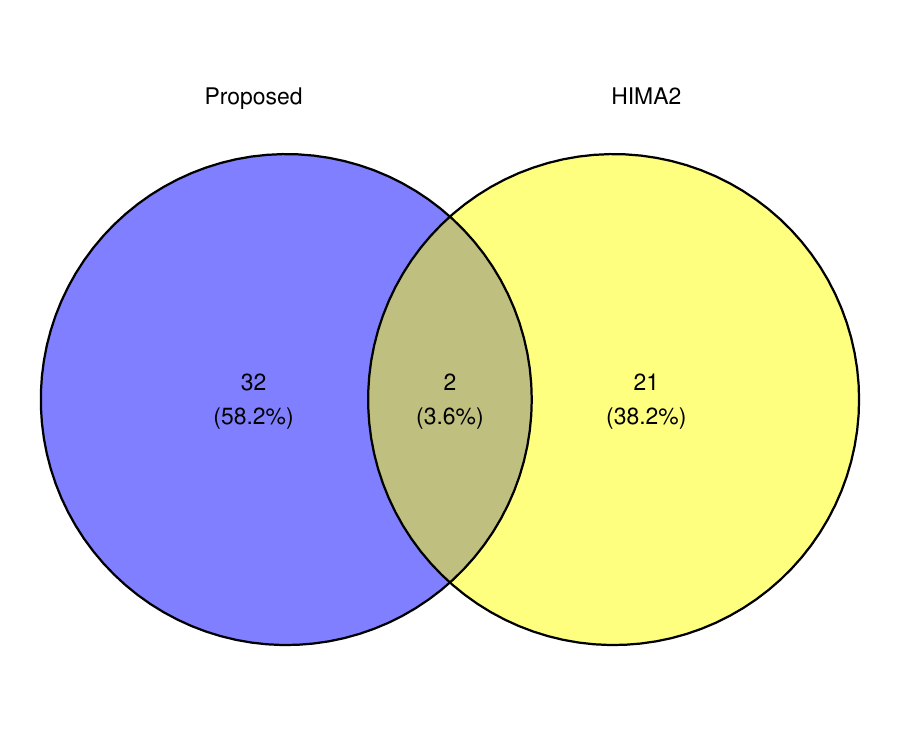}
\caption{Venn diagram illustrating the overlap of significant mediating genes identified by two methods: CHIMA (denoted as Proposed) and HIMA2, when modeling neuronal cell estimates as the outcome. The diagram highlights both shared and method-specific mediators, providing a comparative assessment of mediator selection consistency across approaches.}
\label{LBPfig2}
\end{figure}

\begin{figure}[t]
\centering
\includegraphics[scale=0.65]{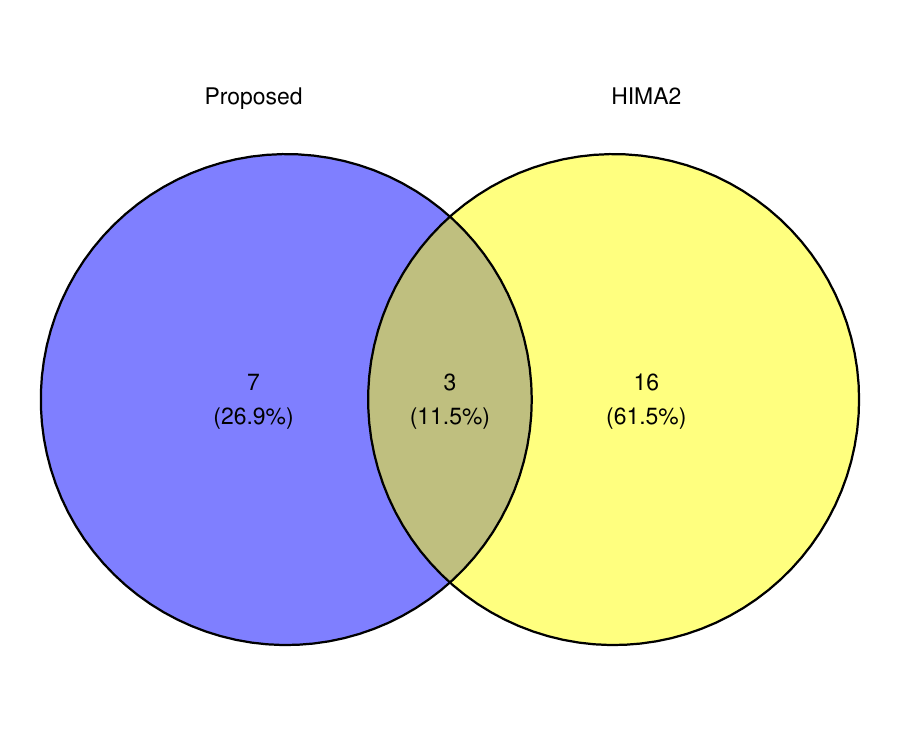}
\caption{Venn diagram illustrating the overlap of significant mediating genes identified by two methods: CHIMA (denoted as Proposed) and HIMA2, when modeling astrocytic gliosis (AST) estimates as the outcome. The diagram highlights both shared and method-specific mediators, providing a comparative assessment of mediator selection consistency across approaches.}
\label{LBPfig3}
\end{figure}

\begin{figure}[H]
\centering
\includegraphics[scale=0.65]{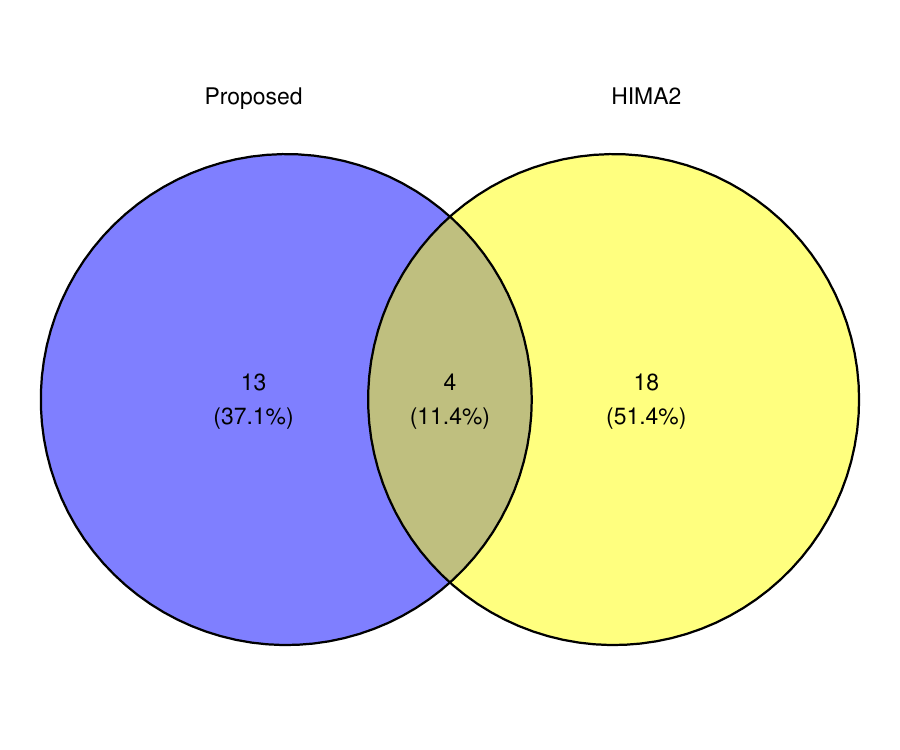}
\caption{Venn diagram illustrating the overlap of significant mediating genes identified by two methods: CHIMA (denoted as Proposed) and HIMA2, when modeling microglial (MG) estimates as the outcome. The diagram highlights both shared and method-specific mediators, providing a comparative assessment of mediator selection consistency across approaches.}
\label{LBPfig4}
\end{figure}

The complete lists of significant mediating genes, with and without covariate adjustment, comparing the proposed method to HIMA2 are provided in Appendix \ref{sec:supp}. Venn diagrams are provided to illustrate the overlap between the two methods (See Figures \ref{LBPfig1} to \ref{LBPfig4}). Notably, the following genes were identified by both methods within each outcome-specific mediation analysis: IFT52, BCL2L11, and ZSCAN30 for ODC; C5orf22 and PCYOX1L for the neuronal outcome; MTMR7, SDS, and YAP1 for AST; and IP6K2, C16orf58, IL16, and TLR5 for MG. As a result of applying the proposed approach, 36 genes for ODC, 34 genes for neuronal, 10 genes for AST, and 17 genes for MG were identified as mediators between PD and each respective cell type. Of the 36 mediator genes identified for ODC, 33 were exclusively detected by the proposed approach. Similarly, among the 34 genes identified for neuronal, 32 were unique to the proposed method. For AST, 7 out of 10 genes were exclusive to the proposed approach, and for MG, 13 of 17 genes were uniquely identified by our method. Overall, these findings highlight the capability of the proposed approach to detect potential mediators and to reveal distinct gene signatures across cell types. 

Among the genes identified by both methods, PCYOX1L, which was detected in the neuronal estimates, has recently been associated with atherosclerotic cardiovascular disease \citep{commongene1}. Similarly, YAP1, identified in the AST estimates, has been implicated in the regulation of adult hippocampal neural stem cells and may contribute to the repression of hippocampal neurogenesis \citep{commongene2}. Interestingly, among the genes exclusively identified by the proposed method, ABCC11, found in the neuronal estimates, carries a non-functional variant that is common in East Asian populations and has been studied in dermatological research related to earwax type and axillary odor \citep{uniquegene1}. In addition, FBX032, uniquely identified in the AST estimates, has been associated with muscle atrophy in previous clinical studies \citep{uniquegene1}. Given that muscle atrophy (or referred to as sarcopenia in aging populations) is a significant concern in individuals with Parkinson's disease and related parkinsonian syndromes, this finding may offer new insight into PD-related mechanisms \citep{sarcopenia}. The analysis was conducted on the Minerva high-performance computing system at the Icahn School of Medicine at Mount Sinai, using 10 CPU cores and 3 GB of RAM per node, and tool approximately 1.476 minutes to complete.

\section{Conclusion}\label{sec:conc}
We proposed CHIMA for high-dimensional mediation analysis to improve performance when mediators are correlated. Its key idea is to apply HOLP for accurate screening and AO for estimation of mediation effects. Our extensive simulation studies demonstrate CHIMA outperforms the existing methods in terms of screening accuracy and power while controlling FDR at nominal level. We also apply CHIMA to detect genes mediating the relationship between Parkinson’s disease status and neural cell-type compositions. Consequently, we could find more significant mediators for ODC and neuronal loss, as well as unique mediators for AST and MG undetected by HIMA2.

For future research, it would be worth to pursue improving the existing approaches for high-dimensional mediation analysis with non-continuous response such as binary or survival outcome. However, a direct extension of CHIMA would be challenging for such non-continuous outcome since both HOLP and AO consider continuous outcome only. Furthermore, as the current model relies on a simple linear effect assumption, it may be overly restrictive in practice. Incorporating more sophisticated models that account for interactions would provide a more comprehensive understanding.

\section{Acknowledgements}
The computing for Section \ref{sec:sim} was performed at the High Performance Computing Center at Oklahoma State University supported in part through the National Science Foundation grant OAC-1531128. The computing for Section \ref{sec:real} was supported in part through the computational and data resources and staff expertise provided by Scientific Computing and Data at the Icahn School of Medicine at Mount Sinai and supported by the Clinical and Translational Science Awards (CTSA) grant UL1TR004419 from the National Center for Advancing Translational Sciences. The last author (S.A.) was supported in part by National Cancer Institute Cancer Center Support Grant P30CA196521–01 awarded to the Tisch Cancer Institute of the Icahn School of Medicine at Mount Sinai and used the Biostatistics Shared Resource Facility. The content is solely the responsibility of the authors and does not necessarily represent the official views of the National Institutes of Health. 

\newpage
\appendix
\numberwithin{equation}{section}
\numberwithin{table}{section}

\newpage
\section{Appendix}\label{sec:app}

\subsection{Details of Step 3}\label{app:step3}
Recall that our goal is to test $H_{0j}:\alpha_{j}\beta_{j}=0$ for each $j\in\mathcal{S}$ where $\mathcal{S}$ denotes the set of candidate mediators identified in Step $1$. To do so, \cite{dai2022multiple} propose a multiple testing method based on the assumption that $p_{\text{max},j}=\text{max}(p_{\alpha,j},p_{\beta,j})$ follows a 3-component mixture distribution under $H_{0j}$ as $H_{0j}$ can be equivalently expressed as the union of the following disjoint null hypotheses:
\begin{equation}\label{compnull}
\begin{aligned}
&H_{00,j}:\alpha_{j}=0\ \text{and}\ \beta_{j}=0,\\    
&H_{01,j}:\alpha_{j}=0\ \text{and}\ \beta_{j}\neq0,\\    
&H_{10,j}:\alpha_{j}\neq0\ \text{and}\ \beta_{j}=0.\\    
\end{aligned}    
\end{equation}
Following \cite{dai2022multiple}, we estimate FDR as
\[
\widehat{\text{FDR}}(t)=\frac{\hat{\pi}_{01}t + \hat{\pi}_{10}t + \hat{\pi}_{00}t^2}{\text{max}[1, R(t)]/|\mathcal{S}|}\quad \text{for}\quad t\in[0,1],
\]
where $\hat{\pi}_{00}$, $\hat{\pi}_{00}$, $\hat{\pi}_{00}$ and $R(t)$ are the estimated proportions of the corresponding null in \eqref{compnull} and the number of rejection at cut-off $t$, respectively. Specifically,  
\[
\begin{aligned}
\hat{\pi}_{01}&= \frac{\sum_{j\in\mathcal{S}}\mathbf{1}\{p_{\alpha,j} > \lambda,p_{\beta,j} > \lambda\}}{(1-\lambda)^{2}|\mathcal{S}|},\\
\hat{\pi}_{10}&= \hat{\pi}_{0+}-\hat{\pi}_{00},\quad
\hat{\pi}_{00}= \hat{\pi}_{+0}-\hat{\pi}_{00} ,\\
\hat{\pi}_{0+}&= \frac{\sum_{j\in\mathcal{S}}\mathbf{1}\{p_{\alpha,j} > \lambda\}}{(1-\lambda)|\mathcal{S}|},\quad
\hat{\pi}_{+0}= \frac{\sum_{j\in\mathcal{S}}\mathbf{1}\{p_{\beta,j} > \lambda\}}{(1-\lambda)|\mathcal{S}|},\\
R(t)&=\sum_{j\in S}\mathbf{1}\{\text{max}(p_{\alpha,j},p_{\beta,j})\leq t\},
\end{aligned}
\]
where $\mathbf{1}\{\cdot\}$ denotes the indicator function and $\lambda\in[0,1]$ is some tuning parameter. Given $\{(p_{\alpha,j},p_{\beta,j})
\}_{j\in\mathcal{S}}$ as its input, the above $\widehat{\text{FDR}}$ can be readily computed by using the \texttt{HDMT} \texttt{R} package. This also implements an automatic selection of the optimal $\lambda$ that yields a more well-behaved fit of normal q-q plot. For more details, we refer to \cite{dai2022multiple} or \url{https://cran.rstudio.com/web/packages/HDMT}.

\newpage
\subsection{Additional simulation results}\label{app:sim}

\textbf{1.} Tables \ref{tab2:screen}-\ref{tab2:fdp}: the results for $p=6000$ under the same settings as in Section \ref{sec:sim}

\begin{table}[H]
\centering
\begin{tabular}{ccccc}
\hline
Covariance structure & $s_{11}$ & \multicolumn{3}{c}{Method} \\
& & HOLP & HIMA2 & HIMA \\
\hline
\multirow{2}{*}{Compound symmetry} & 4 & 0.9915 & 1.0000 & 1.0000 \\
\cline{2-5}
& 6 & 0.9877 & 1.0000 & 1.0000 \\
\hline
\multirow{2}{*}{Toeplitz ($\rho=0.85$)} & 4 & 0.9980 & 1.0000 & 1.0000 \\
\cline{2-5}
& 6 & 1.0000 & 1.0000 & 1.0000 \\
\hline
\multirow{2}{*}{Toeplitz ($\rho=-0.85$)} & 4 & 0.9820 & 0.9780 & 0.7945 \\
\cline{2-5}
& 6 & 0.9760 & 0.9840 & 0.8120 \\
\hline
\multirow{2}{*}{Factor with $r = 2$} & 4 & 0.9590 & 0.9125 & 0.3475 \\
\cline{2-5}
& 6 & 0.9613 & 0.8987 & 0.2980 \\
\hline
\multirow{2}{*}{Factor with $r = 4$} & 4 & 0.9465 & 0.8440 & 0.1975 \\
\cline{2-5}
& 6 & 0.9590 & 0.8380 & 0.1483 \\
\hline
\end{tabular}
\caption{Average screening rate of each method for each different covariance structures and $s_{11}=4,6$ at significance level $\alpha=0.05$ when $n = 400$ and $p=6000$}\label{tab2:screen}
\end{table}

\begin{table}[H]
\centering
\begin{tabular}{ccccccc}
\hline
Covariance structure & $s_{11}$ & \multicolumn{5}{c}{Method} \\
& & CHIMA & HOLP$+$DB & HIMA2 & HIMA & HDMA \\
\hline
\multirow{2}{*}{Compound symmetry} & 4 & 0.7530 & 0.8285 & 0.8000 & 0.7175 & 0.8000 \\
\cline{2-7}
& 6 & 0.7553 & 0.8130 & 0.7790 & 0.7383 & 0.7963 \\
\hline
\multirow{2}{*}{Toeplitz ($\rho=0.85$)} & 4 & 0.8725 & 0.9600 & 0.9510 & 0.7790 & 0.9845 \\
\cline{2-7}
& 6 & 0.8653 & 0.9663 & 0.9640 & 0.7747 & 0.9890 \\
\hline
\multirow{2}{*}{Toeplitz ($\rho=-0.85$)} & 4 & 0.7675 & 0.7170 & 0.6750 & 0.4240 & 0.6250  \\
\cline{2-7}
& 6 & 0.7820 & 0.7280 & 0.6657 & 0.4343 & 0.6367 \\
\hline
\multirow{2}{*}{Factor with $r = 2$} & 4 & 0.9190 & 0.9295 & 0.8685 & 0.3355 & 0.3455 \\
\cline{2-7}
& 6 & 0.9223 & 0.9253 & 0.8403 & 0.2847 & 0.2923 \\
\hline
\multirow{2}{*}{Factor with $r = 4$} & 4 & 0.8960 & 0.9060 & 0.7790 & 0.1855 & 0.1935 \\
\cline{2-7}
& 6 & 0.8970 & 0.9053 & 0.7573 & 0.1363 & 0.1437 \\
\hline
\end{tabular}
\caption{Average power of each method for each different covariance structures and $s_{11}=4,6$ at significance level $\alpha=0.05$ when $n = 400$ and $p=6000$}\label{tab2:power}
\end{table}

\begin{table}[H]
\centering
\begin{tabular}{ccccccc}
\hline
Covariance structure & $s_{11}$ & \multicolumn{5}{c}{Method} \\
& & CHIMA & HOLP$+$DB & HIMA2 & HIMA & HDMA \\
\hline
\multirow{2}{*}{Compound symmetry} & 4 & 0.0335 & 0.0445 & 0.0036 & 0.0192 & 0.1570 \\
\cline{2-7}
& 6 & 0.0320 & 0.0506 & 0.0047 & 0.0094 & 0.1286 \\
\hline
\multirow{2}{*}{Toeplitz ($\rho=0.85$)} & 4 & 0.0443 & 0.0873  & 0.0690 & 0.0177 & 0.1654 \\
\cline{2-7}
& 6 & 0.0426 & 0.0659 & 0.0474 & 0.0140 & 0.1230 \\
\hline
\multirow{2}{*}{Toeplitz ($\rho=-0.85$)} & 4 & 0.0620 & 0.0956 & 0.0570 & 0.0192 & 0.3137 \\
\cline{2-7}
& 6 & 0.0591 & 0.1184 & 0.0796 & 0.0499 & 0.2586 \\
\hline
\multirow{2}{*}{Factor with $r = 2$} & 4 & 0.0511 & 0.0812 & 0.0078 & 0.0177 & 0.0933 \\
\cline{2-7}
& 6 & 0.0509 & 0.0843 & 0.0103 & 0.0181 & 0.0885 \\
\hline
\multirow{2}{*}{Factor with $r = 4$} & 4 & 0.0519 & 0.0901 & 0.0130 & 0.0203 & 0.1314 \\
\cline{2-7}
& 6 & 0.0612 & 0.0975 & 0.0096 & 0.0172 & 0.1439 \\
\hline
\end{tabular}
\caption{Average FDP of each method for each different covariance structures and $s_{11}=4,6$ at significance level $\alpha=0.05$ when $n = 400$ and $p=6000$}\label{tab2:fdp}
\end{table}

\newpage
\noindent\textbf{2.} Tables \ref{tab3:screen}-\ref{tab4:fdp}: the results for $p=6000,8000$ under the same settings as in Section \ref{sec:sim} except non-zero $\alpha_{j},\beta_{j}\sim\text{Unif}(0.5,1.0)$

\begin{table}[H]
\centering
\begin{tabular}{ccccc}
\hline
Covariance structure & $s_{11}$ & \multicolumn{3}{c}{Method} \\
& & HOLP & HIMA2 & HIMA \\
\hline
\multirow{2}{*}{Compound symmetry} & 4 & 0.9905 & 1.0000 & 1.0000 \\
\cline{2-5}
& 6 & 0.9943 & 1.0000 & 1.0000 \\
\hline
\multirow{2}{*}{Toeplitz ($\rho=0.85$)} & 4 & 0.9980 & 1.0000 & 1.0000 \\
\cline{2-5}
& 6 & 0.9983 & 1.0000 & 1.0000 \\
\hline
\multirow{2}{*}{Toeplitz ($\rho=-0.85$)} & 4 & 0.9905 & 0.9700 & 0.7505 \\
\cline{2-5}
& 6 & 0.9873 & 0.9863 & 0.7937 \\
\hline
\multirow{2}{*}{Factor with $r = 2$} & 4 & 0.9675 & 0.9290 & 0.3790 \\
\cline{2-5}
& 6 & 0.9770 & 0.9260 & 0.3300 \\
\hline
\multirow{2}{*}{Factor with $r = 4$} & 4 & 0.9625 & 0.8835 & 0.2095 \\
\cline{2-5}
& 6 & 0.9633 & 0.8717 & 0.1633 \\
\hline
\end{tabular}
\caption{Average screening rate of each method for each different covariance structures and $s_{11}=4,6$ at significance level $\alpha=0.05$ when $n = 400$ and $p=6000$}\label{tab3:screen}
\end{table}

\begin{table}[H]
\centering
\begin{tabular}{ccccccc}
\hline
Covariance structure & $s_{11}$ & \multicolumn{5}{c}{Method} \\
& & CHIMA & HOLP$+$DB & HIMA2 & HIMA & HDMA \\
\hline
\multirow{2}{*}{Compound symmetry} & 4 & 0.8800 & 0.9325 & 0.9345 & 0.8860 & 0.9410 \\
\cline{2-7}
& 6 & 0.8880 & 0.9220 & 0.9100 & 0.8940 & 0.9317 \\
\hline
\multirow{2}{*}{Toeplitz ($\rho=0.85$)} & 4 & 0.9700 & 0.9905 & 0.9925 & 0.8820 & 0.9995 \\
\cline{2-7}
& 6 & 0.9650 & 0.9937 & 0.9960 & 0.8847 & 1.0000 \\
\hline
\multirow{2}{*}{Toeplitz ($\rho=-0.85$)} & 4 & 0.9080 & 0.8495 & 0.7910 & 0.4435 & 0.6370  \\
\cline{2-7}
& 6 & 0.9247 & 0.8630 & 0.8017 & 0.4907 & 0.6737 \\
\hline
\multirow{2}{*}{Factor with $r = 2$} & 4 & 0.9605 & 0.9625 & 0.9250 & 0.3770 & 0.3785 \\
\cline{2-7}
& 6 & 0.9703 & 0.9710 & 0.9180 & 0.3247 & 0.3283 \\
\hline
\multirow{2}{*}{Factor with $r = 4$} & 4 & 0.9565 & 0.9565 & 0.8760 & 0.2075 & 0.2090 \\
\cline{2-7}
& 6 & 0.9523 & 0.9540 & 0.8620 & 0.1600 & 0.1630 \\
\hline
\end{tabular}
\caption{Average power of each method for each different covariance structures and $s_{11}=4,6$ at significance level $\alpha=0.05$ when $n = 400$ and $p=6000$}\label{tab3:power}
\end{table}

\begin{table}[H]
\centering
\begin{tabular}{ccccccc}
\hline
Covariance structure & $s_{11}$ & \multicolumn{5}{c}{Method} \\
& & CHIMA & HOLP$+$DB & HIMA2 & HIMA & HDMA \\
\hline
\multirow{2}{*}{Compound symmetry} & 4 & 0.0302 & 0.0395 & 0.0034 & 0.0130 & 0.1337 \\
\cline{2-7}
& 6 & 0.0282 & 0.0362 & 0.0056 & 0.0070 & 0.1154 \\
\hline
\multirow{2}{*}{Toeplitz ($\rho=0.85$)} & 4 & 0.0445 & 0.0851  & 0.0605 & 0.0092 & 0.1568 \\
\cline{2-7}
& 6 & 0.0411 & 0.0664 & 0.0468 & 0.0066 & 0.1201 \\
\hline
\multirow{2}{*}{Toeplitz ($\rho=-0.85$)} & 4 & 0.0495 & 0.0803 & 0.0564 & 0.0273 & 0.3097 \\
\cline{2-7}
& 6 & 0.0614 & 0.1164 & 0.0771 & 0.0482 & 0.2528 \\
\hline
\multirow{2}{*}{Factor with $r = 2$} & 4 & 0.0493 & 0.0828 & 0.0127 & 0.0132 & 0.0891 \\
\cline{2-7}
& 6 & 0.0423 & 0.0853 & 0.0130 & 0.0197 & 0.0856 \\
\hline
\multirow{2}{*}{Factor with $r = 4$} & 4 & 0.0493 & 0.0853 & 0.0161 & 0.0143 & 0.1217 \\
\cline{2-7}
& 6 & 0.0603 & 0.0949 & 0.0118 & 0.0272 & 0.1381 \\
\hline
\end{tabular}
\caption{Average FDP of each method for each different covariance structures and $s_{11}=4,6$ at significance level $\alpha=0.05$ when $n = 400$ and $p=6000$}\label{tab3:fdp}
\end{table}

\begin{table}[t]
\centering
\begin{tabular}{ccccc}
\hline
Covariance structure & $s_{11}$ & \multicolumn{3}{c}{Method} \\
& & HOLP & HIMA2 & HIMA \\
\hline
\multirow{2}{*}{Compound symmetry} & 4 & 0.9975 & 1.0000 & 1.0000 \\
\cline{2-5}
& 6 & 0.9917 & 1.0000 & 1.0000 \\
\hline
\multirow{2}{*}{Toeplitz ($\rho=0.85$)} & 4 & 0.9995 & 1.0000 & 1.0000 \\
\cline{2-5}
& 6 & 0.9987 & 1.0000 & 1.0000 \\
\hline
\multirow{2}{*}{Toeplitz ($\rho=-0.85$)} & 4 & 0.9910 & 0.9735 & 0.7295 \\
\cline{2-5}
& 6 & 0.9857 & 0.9820 & 0.7763 \\
\hline
\multirow{2}{*}{Factor with $r = 2$} & 4 & 0.9625 & 0.9245 & 0.3705 \\
\cline{2-5}
& 6 & 0.9673 & 0.9177 & 0.3217 \\
\hline
\multirow{2}{*}{Factor with $r = 4$} & 4 & 0.9575 & 0.8750 & 0.1990 \\
\cline{2-5}
& 6 & 0.9600 & 0.8663 & 0.1510 \\
\hline
\end{tabular}
\caption{Average screening rate of each method for each different covariance structures and $s_{11}=4,6$ at significance level $\alpha=0.05$ when $n = 400$ and $p=8000$}\label{tab4:screen}
\end{table}

\begin{table}[H]
\centering
\begin{tabular}{ccccccc}
\hline
Covariance structure & $s_{11}$ & \multicolumn{5}{c}{Method} \\
& & CHIMA & HOLP$+$DB & HIMA2 & HIMA & HDMA \\
\hline
\multirow{2}{*}{Compound symmetry} & 4 & 0.8760 & 0.9275 & 0.9320 & 0.8735 & 0.9270 \\
\cline{2-7}
& 6 & 0.8920 & 0.9283 & 0.9150 & 0.8863 & 0.9340 \\
\hline
\multirow{2}{*}{Toeplitz ($\rho=0.85$)} & 4 & 0.9755 & 0.9955 & 0.9970 & 0.8795 & 1.0000 \\
\cline{2-7}
& 6 & 0.9753 & 0.9947 & 0.9943 & 0.8703 & 0.9993 \\
\hline
\multirow{2}{*}{Toeplitz ($\rho=-0.85$)} & 4 & 0.9105 & 0.8450 & 0.7900 & 0.3925 & 0.6085  \\
\cline{2-7}
& 6 & 0.9223 & 0.8610 & 0.7913 & 0.4700 & 0.6483 \\
\hline
\multirow{2}{*}{Factor with $r = 2$} & 4 & 0.9560 & 0.9580 & 0.9210 & 0.3685 & 0.3700 \\
\cline{2-7}
& 6 & 0.9570 & 0.9603 & 0.9090 & 0.3160 & 0.3200 \\
\hline
\multirow{2}{*}{Factor with $r = 4$} & 4 & 0.9525 & 0.9510 & 0.8650 & 0.1960 & 0.1090 \\
\cline{2-7}
& 6 & 0.9490 & 0.9497 & 0.8537 & 0.1480 & 0.1503 \\
\hline
\end{tabular}
\caption{Average power of each method for each different covariance structures and $s_{11}=4,6$ at significance level $\alpha=0.05$ when $n = 400$ and $p=8000$}\label{tab4:power}
\end{table}

\begin{table}[H]
\centering
\begin{tabular}{ccccccc}
\hline
Covariance structure & $s_{11}$ & \multicolumn{5}{c}{Method} \\
& & CHIMA & HOLP$+$DB & HIMA2 & HIMA & HDMA \\
\hline
\multirow{2}{*}{Compound symmetry} & 4 & 0.0198 & 0.0254 & 0.0053 & 0.0170 & 0.1482 \\
\cline{2-7}
& 6 & 0.0351 & 0.0415 & 0.0055 & 0.0078 & 0.1186 \\
\hline
\multirow{2}{*}{Toeplitz ($\rho=0.85$)} & 4 & 0.0540 & 0.0943 & 0.0733 & 0.0114 & 0.1493 \\
\cline{2-7}
& 6 & 0.0448 & 0.0776 & 0.0578 & 0.0122 & 0.1209 \\
\hline
\multirow{2}{*}{Toeplitz ($\rho=-0.85$)} & 4 & 0.0531 & 0.0807 & 0.0617 & 0.0358 & 0.3367 \\
\cline{2-7}
& 6 & 0.0754 & 0.1300 & 0.0777 & 0.0439 & 0.2610 \\
\hline
\multirow{2}{*}{Factor with $r = 2$} & 4 & 0.0570 & 0.0892 & 0.0161 & 0.0252 & 0.0865 \\
\cline{2-7}
& 6 & 0.0548 & 0.0956 & 0.0158 & 0.0215 & 0.0913 \\
\hline
\multirow{2}{*}{Factor with $r = 4$} & 4 & 0.0659 & 0.0984 & 0.0151 & 0.0203 & 0.1233 \\
\cline{2-7}
& 6 & 0.0576 & 0.1038 & 0.0120 & 0.0155 & 0.1293 \\
\hline
\end{tabular}
\caption{Average FDP of each method for each different covariance structures and $s_{11}=4,6$ at significance level $\alpha=0.05$ when $n = 400$ and $p=8000$}\label{tab4:fdp}
\end{table}

\subsection{Supplementary tables}\label{sec:supp}
\begin{table}[H]
\centering
\setlength{\tabcolsep}{15pt} 
\begin{tabular}{llll}
\toprule
\multicolumn{2}{c}{ Without Covariates} & \multicolumn{2}{c}{ With Covariates*} \\\midrule
\multicolumn{1}{c}{ CHIMA } & \multicolumn{1}{c}{ HIMA2 } & \multicolumn{1}{c}{ CHIMA } & \multicolumn{1}{c}{ HIMA2 } \\
\toprule
ELMO2         & AGK         & ETV1          & ROS1 \\
\textbf{ADCY2} & NUDCD3      & FAM13B        & SSH1 \\
NALCN         & \textbf{ADCY2} & ZC3H11A       & CIRBP \\
\textbf{DDX25} & GNB1        & PAG1          & \textbf{IFT52} \\
\textbf{CCND2} & BRINP1      & CERS4         & JHY \\
NUDT10        & NECAP1      & RAPGEF4       & KCNC1 \\
FIBCD1        & CFAP263     & CINP & BMAL1 \\
TBC1D8B       & MTPN        & \textbf{IFT52} & DPH6 \\
TFCP2         & ANKMY2      & SPAG1         & DYNC2LI1 \\
QDPR          & TNFAIP1     & ENOX1         & TPM3 \\
STARD9        & TBC1D9      & ABCC11        & GPM6A \\
CTSB          & \textbf{DDX25} & NPY           & \textbf{BCL2L11} \\
DIRAS2        & FBXO9       & RAPGEF5       & C2CD2 \\
ANKS6         & \textbf{CCND2} & STX17         & NPR2 \\
SEMA3E        & ITFG1       & BPHL          & CIP2A \\
FAM53B        & SCN2A       & RNF144A       & NIPBL \\
              & ARHGAP20    & \textbf{BCL2L11} & GSTA4 \\
              & GLYR1       & PALM2-AKAP2   & PLEKHM3 \\
              & ZNF215      & ABCG1         & \textbf{ZSCAN30} \\
              & GPM6A       & AKAP13        & DAPK1 \\
              & CAMKV       & FAM86JP       & HMGN1 \\
              & SVOP        & MALT1         & AL023284.4 \\
              & LRRC20      & PEAK1         & AC105285.1 \\               
              & NUDT4       & EID2B         & LINC01844 \\               
              & GALNT11     & \textbf{ZSCAN30}   &  \\           
              & RGS7        & LHFPL3        & \\              
              & PIP4K2B     & ZNF470        & \\
              & LUZP6        & ZNF433        & \\
              & AL592490.1   & SMURF1        & \\
              &             & CPLX3         & \\
              &              & LINC00863  AL512422.1   & \\
              &            & HIST1H2BN  AC097376.1 & \\    
              &              & ZBED6    RPL23AP97  & \\         
\bottomrule
\end{tabular}
\label{tab:odc_tab}
\caption{Table of significant genes found by CHIMA and HIMA2 for ODC changes}
\end{table}

\begin{table}[H]
\centering
\setlength{\tabcolsep}{15pt}
\begin{tabular}{llll}
\toprule
\multicolumn{2}{c}{ Without Covariates} & \multicolumn{2}{c}{ With Covariates*} \\\midrule
\multicolumn{1}{c}{ Proposed } & \multicolumn{1}{c}{ HIMA2 } & \multicolumn{1}{c}{ Proposed } & \multicolumn{1}{c}{ HIMA2 } \\
\toprule
ATP6AP1     & BRINP1      & RB1CC1     & ROS1 \\
C5orf22     & TNFAIP1     & LETMD1     & \textbf{C5orf22} \\
ATRN        & DDX25 & \textbf{C5orf22} & CIRBP \\
NCALD       & KLHDC3      & YPEL3      & PBLD \\
ZFAND5      & ITFG1       & RAPGEF4    & JHY \\
CCND2 & CHL1      & GSS        & ACTR1B \\
GDA         & SCN2A       & BEX4       & KCNC1 \\
MAS1        & ZNF215      & GSDME      & BMAL1 \\
FIBCD1      & GPR158      & ABCC11     & DPH6 \\
TOR1A       & BRINP3      & PMS2       & DYNC2LI1 \\
GLCE        & CAMKV       & ECSIT      & FAM117B \\
C16orf58    & FRMPD4      & BEX2       & FGF5 \\
PAK1        & NUDT4       & TEX15      & TPM3 \\
HMGA2       & RGS7        & GLCE       & \textbf{PCYOX1L} \\
CAMK4       & PCDHA3      & C16orf58   & ARSK \\
YWHAZ       & BAIAP2-AS1  & \textbf{PCYOX1L} & SVOP \\
DIRAS2      &             & PPP1R2     & HEXIM1 \\
TLN2        &             & DAPK1      & ZNF273 \\
FAM131A     &             & ZNF273     & HMGN1 \\
TRAPPC6B    &             & HMGN1      & ZBED6 \\
NIPSNAP1    &             & ZBED6      & LINC01844 \\
SND1        &             & LHFP3      &         \\
PEG3        &             & OR7E13P    &         \\
CPLX3       &             & TRDC       &         \\
AC091132.5  &             & EEF1DP4    &         \\
GTF2H5      &             & SOGA3      AL512422.1 & \\
AL357054.4  &             & RGL2       &         \\
PNMA8C      &             & ETV5        AC132192.1 & \\
            &             & WAC-AS1    &         \\
            &             & AL355338.1 &         \\
\bottomrule
\end{tabular}
\label{tab:neuronal_tab}
\caption{Table of significant genes found by CHIMA and HIMA2 for neuronal loss}
\end{table}

\begin{table}[H]
\centering
\setlength{\tabcolsep}{23pt}
\begin{tabular}{llll}
\toprule
\multicolumn{2}{c}{ Without Covariates} & \multicolumn{2}{c}{ With Covariates*} \\\midrule
\multicolumn{1}{c}{ Proposed } & \multicolumn{1}{c}{ HIMA2 } & \multicolumn{1}{c}{ Proposed } & \multicolumn{1}{c}{ HIMA2 } \\
\toprule
MED29        & MTMR7        & \textbf{MTMR7}   & \textbf{MTMR7} \\
ARHGEF3      & VTA1         & HEBP2           & FNIP2 \\
NPY          & CFAP263      & ACTR5           & RFX2 \\
YAP1         & DNAJC6       & ENOX1           & GSS \\
DUSP6        & RLIM         & \textbf{SDS}    & BCAT2 \\
ASCL1        & ACLY         & MMD2            & TSNAX \\
C16orf58     & SYT11        & \textbf{YAP1}   & PLCG1 \\
\textbf{APP} & \textbf{APP} & C16orf58        & ARPP19 \\
NTRK2        & CYRIB        & FBXO32          & LPIN1 \\
GLUD1        & FAM131B      & ZNF442          & BTF3L4 \\
TNIK         & WDR82        &                 & \textbf{SDS} \\
CATIP        & RANBP3L      &                 & \textbf{YAP1} \\
AHCYL1       & YWHAZ        &                 & OGT \\
ZNF680       & NKAIN3       &                 & GLUD1 \\
PTAR1        & CYRIA        &                 & PLCB3 \\
DEFA3        & ATL1         &                 & PLEKHM3 \\
LINC01896    &              &                 & C1orf61 \\
AC074194.2   &              &                 & AC008957.3 \\
             &              &                 & AL390879.1 \\
\bottomrule
\end{tabular}
\label{tab: ast_tab}
\caption{Table of significant genes found by CHIMA and HIMA2 for AST}
\end{table}

\begin{table}[H]
\centering
\begin{tabular}{llll}
\toprule
\multicolumn{2}{c}{ Without Covariates} & \multicolumn{2}{c}{ With Covariates*} \\\midrule
\multicolumn{1}{c}{ Proposed } & \multicolumn{1}{c}{ HIMA2 } & \multicolumn{1}{c}{ Proposed } & \multicolumn{1}{c}{ HIMA2 } \\
\toprule
TYRO3         & CD200         & \textbf{IP6K2}     & MBTPS2 \\
NALCN         & CEP41         & DTWD1             & JADE2 \\
FSCN3         & TBC1D9        & TARDBP            & HEBP2 \\
SYT11         & \textbf{CCND2}      & TFCP2             & \textbf{IP6K2} \\
TOR1A         & SERINC3       & \textbf{C16orf58}  & SSH1 \\
APP           & CHL1          & CDC42BPA          & CIRBP \\
CRHBP         & EEIG2         & FAM171A1          & COLGALT1 \\
WNT7A         & WDR82         & PACRGL            & SLC38A2 \\
PALM2-AKAP2   & GRB2          & LGI3              & DDX60 \\
TMEM74        & CMSS1         & \textbf{IL16}     & RTF1 \\
IL16          & CHP1          & MPI               & TUBGCP4 \\
CKAP5         & LRRC8B C17orf102 & \textbf{TLR5}     & DUSP6 \\
MIR551B       & CD2AP         & MICA AC011507.1           & \textbf{C16orf58} \\
CPEB1 AL359915.2   & PCDHA11       & Z93241.1        & ZC3H12C \\
AKAP2              & \textbf{PCDH20} &  RF00003           & SPATA2 \\
RF00003  HIST1H2BF &               & RNU12            & DHX57 \\
SNORA61            &      &              & MARF1 \\
\textbf{PCDH20}    &               &                   & \textbf{IL16} \\
AL358613.3         &             &                   & PPP2R2D \\
                   &               &                   & PLEKHM3 \\
                   &               &                   & \textbf{TLR5} \\
                   &               &                   & TOP1 \\
\bottomrule
\end{tabular}
\label{tab:mg_tab}
\caption{Table of significant genes found by CHIMA and HIMA2 for MG activation}
\end{table}

\newpage
\bibliographystyle{apalike} 
\bibliography{reference}

\end{document}